\documentclass[aps,pre,showpacs, twocolumn]{revtex4}

\usepackage{graphicx}
\usepackage{amssymb}
\usepackage{amsmath}
\usepackage{colordvi}
\usepackage{color}

\newcommand{\PT}{{\cal PT}}

\newcommand{\cL}{{\cal L}}

\newcommand{\tW}{\widetilde{W}}

\newcommand{\tpsi}{\tilde{\psi}}
\newcommand{\tmu}{\tilde{\mu}}

\newcommand{\vep}{\varepsilon}

\begin{document}

\title{Quasi-one-dimensional harmonically trapped quantum droplets}

\author{Dmitry A. Zezyulin}

\affiliation{School of Physics and Engineering, ITMO University, St. Petersburg 197101, Russia}

\date{\today}

\begin{abstract}

We theoretically consider effectively one-dimensional quantum droplets in a symmetric Bose-Bose mixture confined in a parabolic trap. We systematically investigate ground and excited families of localized trapped modes which bifurcate from eigenstates of the quantum harmonic oscillator as the number of particles departs from zero. Families of nonlinear modes have nonmonotonous behavior of chemical potential on the number of particles and    feature bistability regions. Excited states are unstable close to the linear limit, but become stable when the   number of particles is large enough. In the limit of large density, we derive a modified Thomas-Fermi distribution. Smoothly decreasing  the trapping strength down to zero, one can dynamically transform the ground state solution to the solitonlike quantum droplet, while  excited trapped states break in several moving quantum droplets.

\end{abstract}

\maketitle

\section{Introduction}
\label{intro}

Formation of liquidlike  quantum droplets in weakly interacting    Bose-Bose mixtures is a remarkable manifestation   of the beyond-meanfield effects \cite{Petrov2015}.  In   three-dimensional mixtures, the existence of quantum droplets becomes possible due to the presence of quantum fluctuations which stabilize the system against collapse. At the same time, the liquid phase also persists  in low-dimensional geometries \cite{Petrov2016}.   Quantum droplets have been created in several experiments with two-component  mixtures \cite{Cabrera2018,Semeghini2018,Cheiney2018,collisions,hetero} (and, prior to that,   atomic droplets stabilized by quantum fluctuations have been realized in single-component gases of dipolar atoms \cite{Kadau2016,Ferrier-Barbut2016,Chomaz2016,Schmitt2016}).  The beyond-meanfield corrections that  enable   formation of  quantum droplets in a two-component mixture can be taken into account using  a system of two   Gross-Pitaevskii (GP) equations (or using a single equation, in the case of symmetric mixture), whose specific form heavily depends on the effective dimensionality of   liquid \cite{Petrov2015,Petrov2016} and is essentially different from the previously studied GP   equations   with  cubic or cubic-quintic nonlinearity \cite{PS2,PS}. Effectively one-dimensional (1D) quantum droplets have been studied in several works  \cite{AstaMalo,Abdullaev,Otajonov,Parisi1,Parisi2,Tylutki,Mithun,Karta2022,Zhou2019}.   In particular, it has been found that these states    feature solitonlike behavior and   rich dynamics \cite{AstaMalo}.  A recent study \cite{Karta2022}  presents the analysis of kinks and holes nestling in the spatially extended binary mixture. Solutions of this type can be interpreted as counterparts of conventional dark solitons \cite{Burger,Kevrekidis}.  Multidimensional  quantum droplets have also been in the focus of  active recent research,  see e.g. \cite{Nilsson2019,Dong2021,vortex,vortex2,Karta2019,39K} and review papers \cite{review2019,review,review2}.

A particularly interesting topic is an effect of external trapping on the properties of quantum droplets. For multidimensional   quantum droplets in dipolar gases confined in a harmonic trap,   it has been found that the resulting ground state phase diagram can feature a region of multistability \cite{bistability}.  Modulational instability in trapped dipolar Bose-Einstein condensates (BECs)  leads to   formation of  multiple droplets \cite{MI}.  For quantum droplets in binary mixtures, annular potentials can  facilitate the formation of rotating multidimensional droplets \cite{Nilsson2019,Dong2021}. Formation and dynamics of quantum droplets of bosonic mixtures loaded in  1D  optical lattices has been studied in \cite{lattice} and \cite{Zhou2019}.  Various aspects related to the effectively  nonlinear behavior  of  quantum droplets, such as   the onset of instabilities, bifurcations of nonlinear states from the linear limit, adiabatic excitation of quantum droplets, and symmetry-breaking have been explored for  potentials of different shapes \cite{Liu,Song}.

In the meanfield theory of BECs, it is well known that, apart from the  fundamental ground state, externally trapped condensates can  also exist in the so-called nonground (or excited) states \cite{Yukalov1,Yukalov2,Konotop}.  The first (single-node) nonground state can be interpreted as a trapped dark soliton  \cite{Kevrekidis}  in the effectively 1D geometry or as a vortex state   \cite{Fetter} in the 2D geometry. Experimental realization of these states can be achieved  using the phase-imprinting method \cite{Burger,Denschlag}. More complex  excited states have     wavefunctions with      incrementally increasing   number  of zeros and can be considered as nonlinear  states of the macroscopic quantum oscillator \cite{KAT}.  Various properties of such trapped excited states have been systematically considered  in numerous publications for  1D   (cigar-shaped) geometry  \cite{KAT,nonlin_homo,Kunze,Rodrigues,cigar-shape,Theo,AlZez,CPK10,Brtka,Carr2001} as well as for  multidimensional cases, see in particular   \cite{Carr1,Carr2,Pu,VP1,VP3,VP4,unharm,Fin,Michalache2006,Malomed2007,RadSymmKevrek2007,Zezyulin2009} and collections of available results in \cite{Fetter,Carr,MalomedReview,Karta2019}. 
The excited   states can be dynamically stable \cite{AlZez,Zezyulin2009,CPK10} and    perform persistent periodic motion   around   center of the trap \cite{Brazhnyi,PeliPRE,Peli}.

Vast body of knowledge accumulated for   trapped BECs naturally suggests to deepen our understanding of the role of  external confinement in the formation and behavior of quantum droplets and, in particular, to explore in a more systematic way  the corresponding   nonground states that can potentially emerge in the presence  of the confinement. In this  paper, we aim to perform a systematic study of one-dimensional quantum droplets in a symmetric Bose-Bose mixture loaded in a harmonic (parabolic) potential. Apart from the ground  nodeless states, the resulting system   admits a sequence of families of excited states whose wavefunctions have the incrementing   number of zeros and bifurcate from the eigenstates of the quantum harmonic oscillator. We demonstrate that in the presence of the trapping either the ground state family and the excited families have  bistability regions, where stable states with different numbers of particles coexist at the same value of the chemical potential. Peculiar spectrum of the quantum harmonic  oscillator results in the   instabilities of small-amplitude quantum droplets from the   excited families. These instabilities, however, disappear as the number of particles increases above a certain threshold. In the large-density limit, the trapped states can be described by a modified Thomas-Fermi approximation.  Numerical simulations of dynamics indicate that smooth decrease of the trapping strength down to zero transforms the ground state to the solitonlike quantum droplet,  \textcolor{black}{and  nonground  states break  into several   quantum droplets   moving with different velocities.}  
We also simulate   periodic motion of the quantum droplets around the center of the trap.  Several similarities and dissimilarities are found between the trapped beyond-meanfield system and the model with conventional cubic interactions, as well as in comparison to  the effectively 1D   model with the beyond-meanfield corrections but  without the trap.

Organization of the paper is as follows. In the next Section~\ref{sec:model} we formulate the governing model equation.  Section~\ref{sec:stationary} presents a detailed study of stationary modes, and  \textcolor{black}{Section~\ref{sec:dyn} addresses several dynamical scenarios corresponding to the found states}. Concluding Section~\ref{sec:Concl} summarizes the main results and briefly outlines possible directions for future work.

\section{Model}
\label{sec:model}

In the effectively 1D   geometry, formation of liquid droplets results from the balance between the meanfield repulsive contribution    to the energy per particle  and a beyond-meanfield  attractive correction. In case of the symmetric mixture of two species ($\uparrow$ and $\downarrow$), the dynamics can be described by a single  modified  GP  equation. \textcolor{black}{Assuming that atoms are harmonically confined in a quasi-1D  (i.e., cigar-shaped) geometry with the transverse trapping frequency  $\omega_\bot$ being much larger that the frequency of the longitudinal trapping $\omega_0$, we   use the following model
\cite{Petrov2016,AstaMalo}
\begin{eqnarray}
\label{GPE-phys}
i\hbar \Psi_t =  - \frac{\hbar^2}{2m} \Psi_{xx} - \frac{\sqrt{2m}}{\pi \hbar} g^{3/2} |\Psi|\Psi + \delta g |\Psi|^2\Psi  \nonumber\\[2mm]+ \frac{m}{2} \omega_0^2 x^2 \Psi,
\end{eqnarray} 
where   $g = g_{\uparrow\uparrow}=g_{\downarrow\downarrow}>0$ is the repulsive intraspecies  coupling coefficient, and   $\delta g =  g_{\uparrow\downarrow} + \sqrt{ g_{\uparrow\uparrow}g_{\downarrow\downarrow}}$ gives the difference between the intracpecies repulsion and intraspecies attraction $g_{\uparrow\downarrow}<0$. We assume $\delta g>0$. In order to present our main results, we transform Eq.~(\ref{GPE-phys}) into the dimensionless form adopting characteristic units and normalization of the wavefunction that  result in equal coefficients in front of the nonlinear terms \cite{AstaMalo}:
\begin{eqnarray}
x = \frac{\pi \hbar^2\sqrt{\delta g}}{{2}m g^{3/2}} x', \quad t_0 = \frac{\pi^2 \hbar^3{\delta g}}{ {2}m g^{3}} t',
\end{eqnarray}
where $x'$ and $t'$ are dimensionless variables, and 
\begin{eqnarray}
\Psi = \frac{\sqrt{2m} g^{3/2}}{\pi \hbar\delta g} \Psi'.
\end{eqnarray}
Rewriting Eq.~(\ref{GPE-phys}) and omitting primes we arrive at the following normalized equation
\begin{eqnarray}
\label{GPE-basic}
i\Psi_t =  -  \Psi_{xx} + \nu^2 x^2 \Psi -  |\Psi| \Psi + |\Psi|^2 \Psi,
\end{eqnarray}
where $\nu = \hbar^3\pi^2\delta g \omega_0 / (4mg^3)$ is a dimensionless coefficient that governs the parabolic trap strength.
}

\textcolor{black}{
In what follows, we analyze  a slightly more general model that explicitly contains nonlinear coefficients $\sigma_2\geq 0$ and $\sigma_3\geq 0$ in front of the nonlinear terms:
\begin{eqnarray}
 \label{GPE}
i\Psi_t =  -  \Psi_{xx} + \nu^2 x^2 \Psi - \sigma_2 |\Psi| \Psi + \sigma_3 |\Psi|^2 \Psi.
\end{eqnarray}
While we bear in mind that the default case corresponds to $\sigma_2=\sigma_3=1$,  the relevance of  Eq.~(\ref{GPE})  is justified   by the fact that the generalization  to  other    values of $\sigma_2$ and $\sigma_3$ is technically simple and yet may yield some additional understandings. }

Temporal dynamics governed by Eq.~(\ref{GPE}) conserves  quantities   $N = \int_{-\infty}^\infty |\Psi|^2 dx$ and
\begin{eqnarray}
E = \int_{-\infty}^\infty \left(  |\Psi_x|^2 +  \nu^2 x^2 |\Psi|^2  - \frac{2\sigma_2}{3} |\Psi|^3  + \frac{\sigma_3}{2}|\Psi|^4\right)dx,
\label{eq:energy}
\end{eqnarray}
\textcolor{black}{which give  the  number of atoms and total energy {for  each component of the mixture}, respectively (with the normalization explained below). }

\section{Stationary modes}
\label{sec:stationary}

\subsection{Families of nonlinear modes}

Stationary nonlinear modes for Eq.~(\ref{GPE})    admit  the   representation $\Psi(x,t) = e^{-i \mu   t}\psi(x)$, where $\mu$ is \textcolor{black}{the dimensionless chemical potential}. Spatial shape of the stationary wavefunction  $\psi(x)$ is determined   by the following equation:
\begin{eqnarray}
\label{stat}
\psi_{xx}  + (\mu - \nu^2 x^2)\psi + \sigma_2|\psi|\psi - \sigma_3\psi^3=0,
\end{eqnarray}
subject to the zero boundary conditions at   infinity: $\lim_{x\to \pm \infty} \psi(x)=0$. 
The case $\nu=0$ and $\sigma_2=\sigma_3=1$ was in detail  analyzed in \cite{Petrov2016,AstaMalo}. In this case an explicit solitonlike solution is available which  has the form   $\psi_s(x) = -3\mu  [ 1+ \sqrt{1+9\mu/2} \cosh \sqrt{-\mu x^2}]^{-1}$. It   exists within the finite interval of chemical potentials   $\mu \in (-2/9, 0)$, such that $\lim_{N\to0^+} \mu(N) = 0$ and  $\lim_{N\to\infty}\mu(N) = -2/9$.

In the linear case $\sigma_2=\sigma_3=0$, Eq.~(\ref{stat}) transforms to an eigenvalue problem whose spectrum is well-known. It consists of a sequence of equidistantly spaced discrete  eigenvalues which can be listed in the ascending order as $\tmu_n = \nu( 2n+1)$, where index $n=0,1,\ldots$  enumerates the eigenstates.  The corresponding eigenfunctions   $\tpsi_n(x)$ read
\begin{eqnarray}
\label{eq:twn}
\tpsi_n(x) = \left. {\sqrt[4]{\nu}} H_n(\sqrt{\nu} x)e^{-\nu x^2/2} \right / \sqrt{\sqrt{\pi}2^nn!},
\end{eqnarray}
where $H_n(x)$ are  Hermite  polynomials \cite{AS}. Equation (\ref{eq:twn}) implies the     normalization  $\int_{-\infty}^\infty \tpsi_n^2dx=1$. (Notice that hereafter  we use tildes to distinguish the solutions that  pertain to the linear case).

Regarding the nonlinear stationary equation (\ref{stat}),   for the case of   meanfield nonlinearity ($\sigma_2=0$) it is rather well-known  \cite{KAT,Kunze,Rodrigues,AlZez}  that families of nonlinear modes branch off from the trivial zero solution $\psi(x)\equiv 0$ at $\mu = \tmu_n$. To designate the corresponding bifurcation, we will sometimes say that nonlinear modes bifurcate from  \emph{the linear limit}. Looking for a  similar bifurcation for trapped quantum droplets described by Eq.~(\ref{stat})   with $\sigma_{2}\ne 0$, we use the following perturbation expansions for  small-amplitude nonlinear modes:
 \begin{eqnarray}
 \label{eq:expans1}
\psi_n(x)=\vep\tpsi_n + \vep^2 \psi_n^{(2)} + \vep^3 \psi_n^{(3)} + \ldots,\\
\label{eq:expans2}
 \mu_n = \tmu_n + \vep \mu_n^{(1)}  +
\vep^2\mu_n^{(2)} +\ldots,
\end{eqnarray}
where  $\vep\ll 1$ is a   small real parameter whose  meaning is evident: close to the bifurcation, for the number of particles  corresponding to $\psi_n$   we have $N_n =  \int_{-\infty}^\infty \psi_n^2dx  =  \vep^2 + o(\vep^2)$. Proceeding in the standard way, we substitute expansions (\ref{eq:expans1})--(\ref{eq:expans2}) to Eq.~(\ref{stat}) and collect the terms having equal powers of $\vep$.  While at the order $\vep$ the resulting equation is  satisfied automatically, at  the order $\vep^2$ we obtain $(\partial_x^2 + \tmu_n - \nu^2 x^2) \textcolor{black}{ \psi_n^{(2)}} = -\textcolor{black}{\mu_n^{(1)}} \tpsi_n - \sigma_2 |\tpsi_n|\tpsi_n$. The solvability condition for the latter equation requires its right-hand side to be orthogonal
to $\tpsi_n$. This   requirement determines the leading correction to the  chemical potential:
\begin{equation}
\mu_n^{(1)} = - \sigma_2\int_{-\infty}^\infty \tpsi_n^2 |\tpsi_n| dx.
\end{equation}
The latter coefficient is obviously nonzero, and,  close to the bifurcation point,  the dependence of the nonlinearity-induced shift of chemical potential   on the number of particles $N_n$ is    nearly square-root: 
$\mu - \tmu_n \approx \mu_n^{(1)} \sqrt{N_n}$,
which is in contrast to the linear law $|\mu-\tmu_n| \propto N_n$ in  the case of cubic interactions   and the power law $|\mu|\propto N^{2/3}$ for 1D quantum droplets without the trapping potential \cite{AstaMalo}.  For $\sigma_2>0$ the coefficient $\mu_n^{(1)}$ is obviously negative, which means that sufficiently close to the bifurcation, i.e., for $0<N_n\ll 1$,      chemical potential  of the nonlinear family  $\mu_n$ is less than that of  the linear mode: $\mu_n<\tmu_n$. This   behavior is typical for BECs dominated by the   attractive nonlinearity. However, it can be expected that as the effective nonlinearity becomes stronger,  the  system will be dominated by the cubic repulsive nonlinearity for which the typical behavior is $d\mu/dN>0$. 

\begin{figure}
	\begin{center}
		\includegraphics[width=0.999\columnwidth]{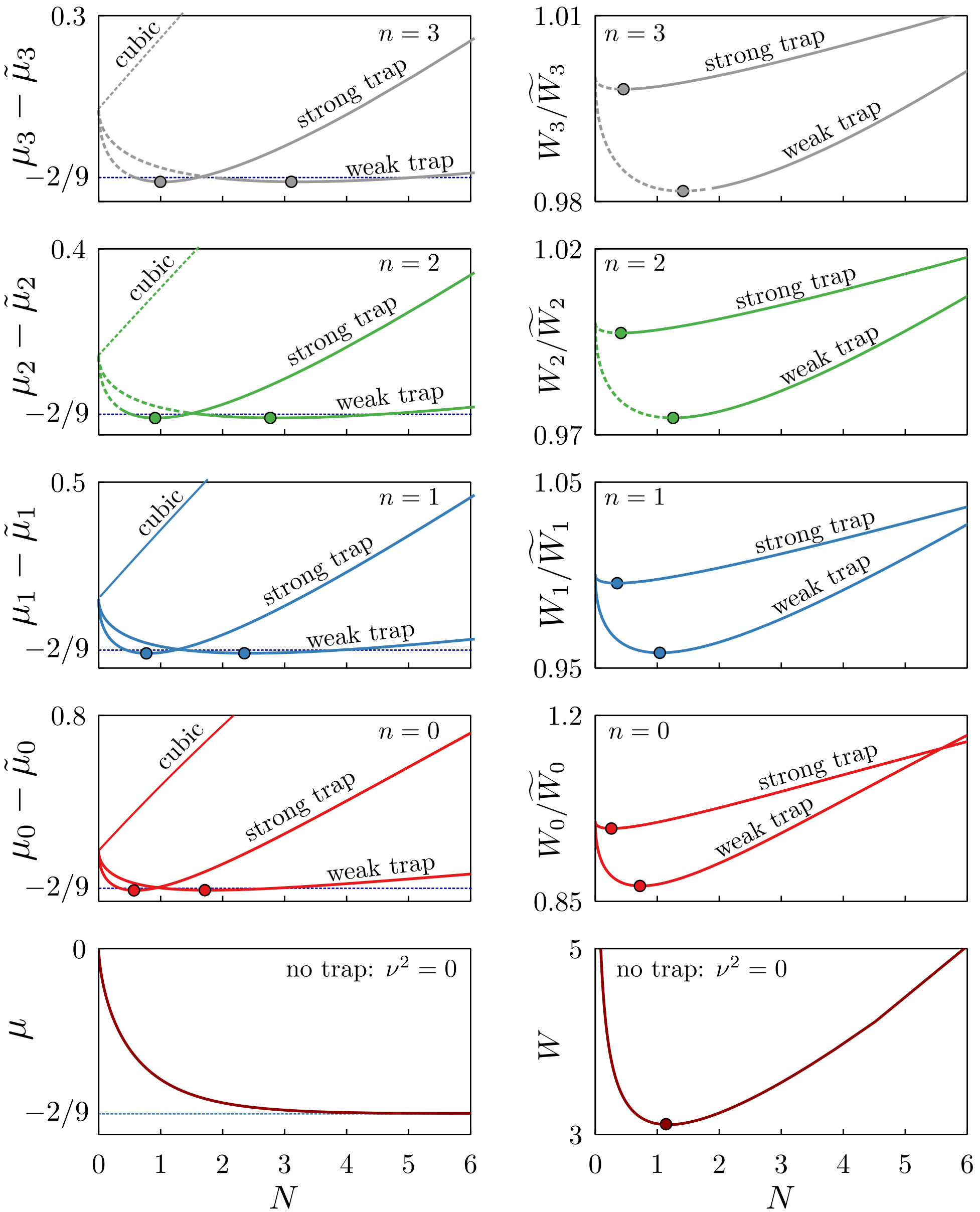}
	\end{center}
	\caption{Dependencies of chemical potential $\mu$ (on the left) and meansquare width $W$ (on the right) on the number of particles $N$ for  the  solitonlike quantum droplet with no trapping  (lower panels, $\nu^2=0$) and for several   families ($n=0,1,2,3$) in the presence of   harmonic trap of two difference strengths: ``strong trap'' $\nu^2=1$ and ``weak trap'' $\nu^2=0.01$. For trapped states, each panel shows the difference $\mu_n - \tmu_n$, where $\tmu_n$ is     the $n$th eigenvalue of the linear problem. For widths of trapped states, we plot  ratios $W_n/\widetilde{W}_n$, where $\tW_n$ is the width of the corresponding  linear eigenfunction $\tpsi_n$. 		 Small circles show   minima of the curves.
	For comparison, in panels with chemical potentials we   plot the analogous dependencies for purely cubic meanfield nonlineaity. For curves labeled as ``cubic'' we have $\sigma_{2}=0$, $\sigma_3=1$ and for all other curves $\sigma_2=\sigma_3=1$.  Solid and dotted fragments of plotted curves correspond to stable and unstable solutions, respectively. This figure shows only the behavior near the linear limit, i.e., for relatively small number of particles $N$.  A ``global'' picture for larger numbers of particles is presented in Fig.~\ref{fig:global}.   }
	\label{fig:families}
\end{figure}

\begin{figure}
	\begin{center}
		\includegraphics[width=0.99\columnwidth]{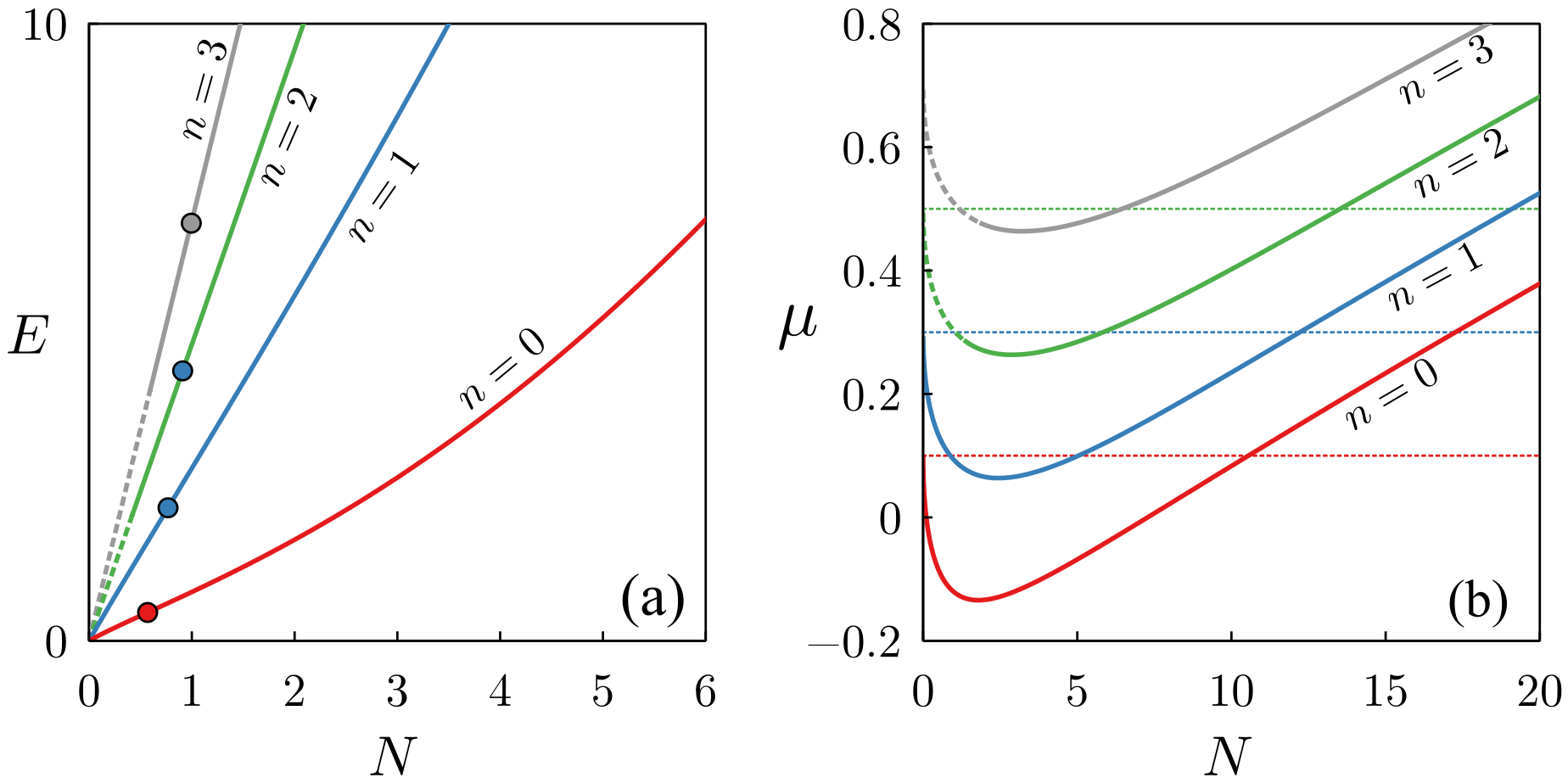}
	\end{center}
	\caption{(a) Dependencies of energy $E$ on  number of particles $N$  for several lower families of nonlinear modes ($n=0,1,2,3$) in the presence of strong  harmonic trap $\nu^2=1$. Circles correspond to the points $N=N^*_n$ of zero curvature, where $\partial^2E(N^*_n)/\partial N^2 = 0$.   (b) Dependencies $\mu(N)$ in the weak trap $\nu^2=0.01$. Thin   horizontal lines correspond to chemical potentials of linear states. In both panels, solid and dotted fragments of plotted curves correspond to stable and unstable solutions, respectively. In this figure $\sigma_2=\sigma_3 = 1$.}
	\label{fig:energy}
\end{figure}

The  nonmonotonous behavior of the chemical potential $\mu$ on number of particles $N$ has indeed been observed for numerically obtained stationary modes, either for the family of ground states and for families of excited states.   In left panels of  Fig.~\ref{fig:families} we plot   families of stationary   modes bifurcating   from four linear eigenstates ($n=0,1,2,3$) and visualized as dependencies $\mu_n(N_n) - \tmu_n$. In our numerical simulations we have considered harmonic trapping of two different strengths: $\nu^2=1$ (``strong trap'') and $\nu^2=0.01$ (``weak trap'').  For comparison, in Fig.~\ref{fig:families} we additionally plot the analogous dependence for quantum droplets with zero trapping $\nu^2=0$, where $\mu(N)$  is a monotonously decreasing function.    As  expected from the above considerations, in the presence of the trapping each dependence $\mu_n(N_n)$ is nonmonotonous and has   a number of particles $N_n^*$,   where the chemical potential acquires its mimimal value $\mu_n(N^*_n) = \mu_n^*$. For each family the difference between the chemical potential of the corresponding linear state $\tmu_n$ and the minimal chemical potential $\mu_{n}^*$ has  approximately the same value: $\Delta_n := \tmu - \mu_n^* \approx  0.24$.  Moreover,   this ``universal'' value does not   change much subject to the change of the trap strength (compare the curves for  strong and weak trapping). The numerical estimate $\Delta_n\approx  0.24$ is   rather close to the analytical value $2/9\approx 0.22$ that limits the existence range of   chemical potentials   in the absence of the    confinement. In the meantime, the  number of particles  $N_n^*$, where the minimal chemical potential $\mu_n^*$ is achieved, is appreciable different for the   considered trap strengths: \textcolor{black}{ for   weak trap the   minimum  of $\mu_n^*$     is achieved at   larger number of particles.} For   fixed trap strength, the sequence of critical numbers of particles $N_n^*$ is increasing:  in particular, for the strong trap we get $N_0^* \approx 0.57$,   $N_1^* \approx 0.77$,    $N_2^* \approx 0.91$, $N_3^* \approx 0.99$. The existence of  a global minimum of chemical potential can have implications for     thermodynamic properties of the condensate. Indeed,   since for families of stationary states we have $\mu_n = \partial{E_n}/\partial N_n$, where $E_n$ is the energy defined by (\ref{eq:energy}),  then around the  point $N^*_n$ the  dependence $E(N)$   has zero curvature: $E(N_n)=E(N^*_n)  + \mu^*_n(N_n-N^*_n) + O((N_n-N_n^*)^3)$. Since for each family the dependence $E_n(N_n)$ is a monotonously increasing function [plotted in Fig.~\ref{fig:energy}(a)], the existence of extrema  $N^*_n$   allows to find pairs of nonlinear states with equal chemical potentials but different energies.

In right panels of Fig.~\ref{fig:families} we show  the meansquare width of nonlinear  modes defined as $W_n = \sqrt{ N_n^{-1} \int_{-\infty}^\infty x^2 |\psi_n|^2 dx}$.  The dependencies $W_n(\mu_n)$ are also nonmonotonous, but the numbers of particles corresponding to the  droplets  of   minimal size     are different from those corresponding the minimal chemical potentials (at the same time, for different families and fixed trap strength the minimal width is achieved for approximately the same number of particles). The nonmonotonous dependence of the meansquare width on the number of particles is similar to that for untrapped solitonlike quantum droplets (see the downmost panel in the right column of Fig.~\ref{fig:families}). However, in contrast to the case of zero trap, in the linear limit $N\to 0$ the widths of trapped states remain finite and do  not diverge.

\begin{figure}
	\begin{center}
		\includegraphics[width=0.99\columnwidth]{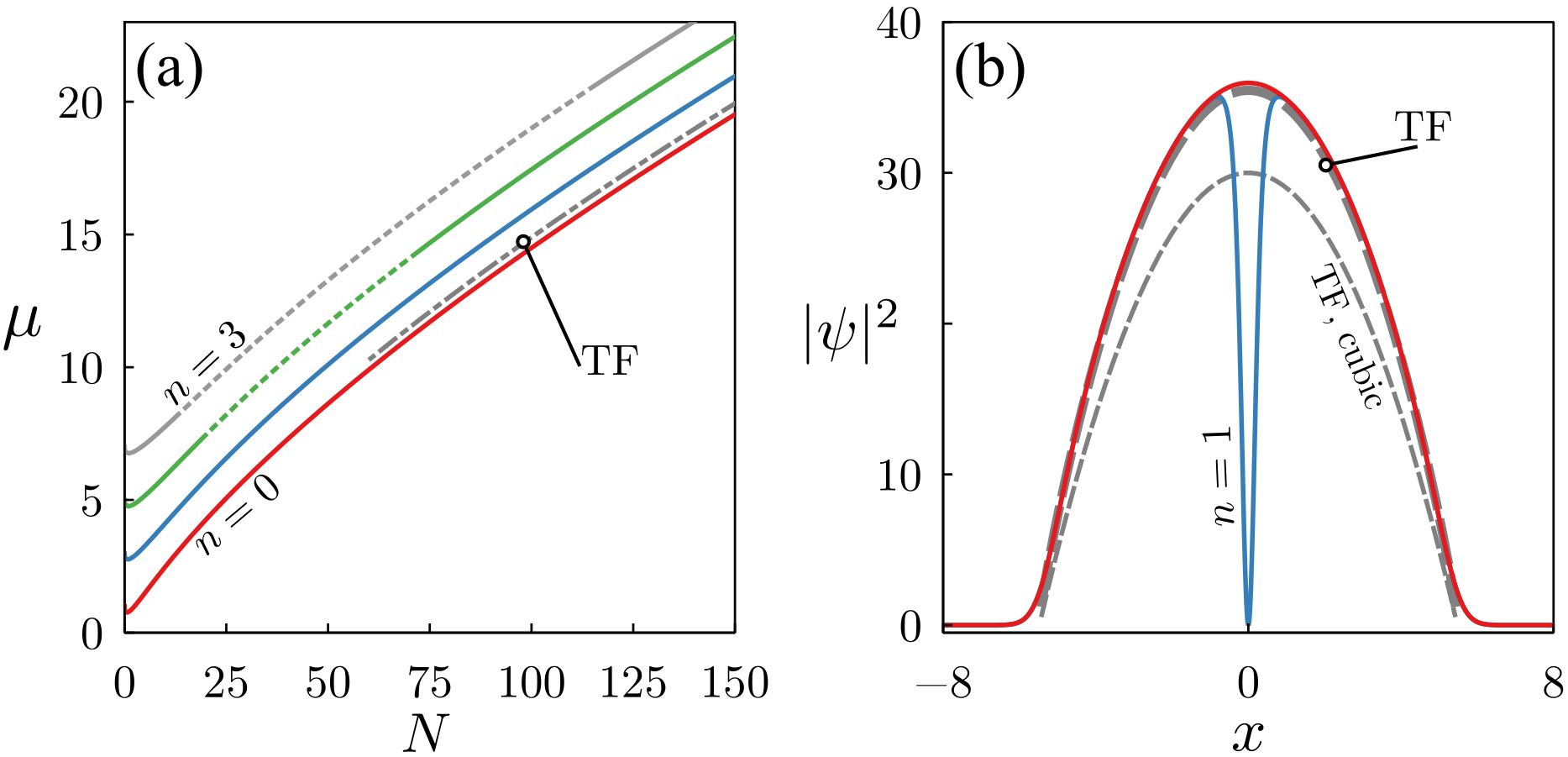}
	\end{center}
	\caption{(a) Dependencies of chemical potential $\mu$ on number of particles $N$  for several   families of nonlinear modes ($n=0,1,2,3$) in the presence of the harmonic trap. Solid and dotted fragments correspond to stable and unstable solutions, respectively. Dash-dotted line labelled as `TF' shows the dependence (\ref{eq:TFN}) obtained analytically for the  ground state family in the Thomas-Fermi limit. (b) Nonlinear modes for $n=0$ and $n=1$ at $\mu=30$. Bold dashed line labelled as `TF' shows the analytical profile in the Thomas-Fermi limit obtained from Eq.~(\ref{eq:TF}). For comparison, with thin dashed line labelled as ``TF cubic'' we show  the  conventional   Thomas-Fermi cloud \cite{PS2} $\mu - \nu^2 x^2$.   In this figure, we consider  strong trap   $\nu^2=1$  and $\sigma_{2}=\sigma_3=1$.}
	\label{fig:global}
\end{figure}

While Fig.~\ref{fig:families} zooms in  the behavior of nonlinear modes  close to the  linear limit, in Fig.~\ref{fig:global}(a) we present a more global diagram which shows the behavior of nonlinear modes in the region of strong effective nonlinearity. In the limit $\mu\gg 1$ and $N\gg 1$ (i.e., the  Thomas-Fermi (TF) limit \cite{PS2,PS}), the  large-density modes are dominated by the cubic nonlinearity. Using $\mu^{-1/2}$ as a small parameter, we observe that in the leading order the TF distribution of the ground state family coincides with the standard one \cite{PS2}, i.e., $\psi^2_{0, TF} =  \sigma_3^{-1}( {\mu - \nu^2 x^2})$ for $x$ lying within the TF radius: $|x|\leq  \nu^{-1}\sqrt{\mu}  $.  Taking into account the next order of the asymptotic series, we find that the beyond-meanfield correction leads to the following modification of the TF ground state solution:
\begin{equation}
\label{eq:TF}
\psi^2_{0, TF} \approx \sigma_3^{-1}(   {\mu - \nu^2 x^2} + {\sigma_2} \sqrt{\mu - \nu^2 x^2} / \sqrt{\sigma_3}),
\end{equation}
which is illustrated in Fig.~\ref{fig:global}(b). The   number of particles corresponding to  the modified TF distribution (\ref{eq:TF}) amounts to  
\begin{equation}
\label{eq:TFN}
N_{0, TF} = ( \sigma_3\nu)^{-1} \left(\frac{4}{3} \mu^{3/2} + \frac{\pi\sigma_2}{2\sqrt{\sigma_3} } \mu\right).
\end{equation}
Therefore, although the beyond-meanfield correction does not change the TF radius, it  results in the positive (and linear in $\mu$) addition to the number of particles (which might seem counterintuitive in view of the fact that the nonlinear terms proportional to $\sigma_2$ and $\sigma_3$ are competing).

\subsection{Bistability of trapped states}

Let us now proceed to discussion of  stability of stationary nonlinear modes. Standard procedure of linear stability analysis (see Appendix~\ref{app:stab}) indicates that dynamical   behavior of small-amplitude perturbations on top of the nonlinear mode is determined by the spectrum of the following   eigenvalue problem:
\begin{equation}
\label{eq:eig}
\Lambda \zeta = L^+L^-\zeta,
\end{equation}
where 
\begin{equation}
\label{eq:Lm}
L^\pm = \partial_x^2 + \mu - \nu^2 x^2 + \frac{\sigma_2}{2}(3\pm 1)|\psi|  - \sigma_3 (2\pm 1) \psi^2,
\end{equation}
$\Lambda$ is the eigenvalue, and $\zeta = \zeta(x)$ is the corresponding eigenfunction.  Stationary mode $\psi(x)$ is said to be stable if all eigenvalues $\Lambda$ are real and nonnegative. Otherwise the solution $\psi(x)$ is said to be unstable, and the growth rate of the exponential dynamical instability is   determined by the positive imaginary part Im$\sqrt{\Lambda}$.  Eigenvalue problem (\ref{eq:eig}) has two evident analytic solutions. The first one corresponds to $\Lambda=0$ with \textcolor{black}{ eigenfunction}  $\zeta = \psi(x)$ and obviously reflects the invariance of the model under the phase rotation. The second analytical solution (which is the peculiarity of   the parabolic potential) is given as $\Lambda=4\nu^2$ and $\zeta = x\psi(x)$ \cite{AlZez,Peli} and proves to be useful for understanding of linear stability of small-amplitude nonlinear modes.

The equidistant spectrum of the parabolic potential   results in the specific stability picture of small-amplitude nonlinear states \cite{AlZez}:   for small-amplitude modes bifurcating from the $n$th linear state, the stability spectrum  in the linear limit contains exactly  $n$ double eigenvalues that result from  the ``resonances'' between different intrinsic modes. When small-amplitude nonlinear states branch off from the linear limit, each double eigenvalue splits either into a pair of real eigenvalues or into a complex-conjugate pair, and  the latter situation implies that the bifurcating small-amplitude modes are unstable. Splittings of double eigenvalues can be analyzed using the  perturbation theory which was previously used in several similar situations \cite{AlZez,Zezyulin2009,Zez2012,Zez2016,Alfimov2019} and, for self-containment of our paper,   is summarized in Appendix~\ref{app:stab}. The results of the perturbation analysis for small-amplitude modes  can be outlined as follows. For the lowest family, $n=0$, there is no double eigenvalues in the spectrum, and therefore  small-amplitude ground states are stable. For the single-node family, $n=1$, there is exactly one double eigenvalue equal to   $\Lambda=4\nu^2$. Splitting of this double eigenvalue into a complex-conjugate pair is \emph{a priori}  impossible due to the presence of the exact solution mentioned above (because the eigenvalue   $\Lambda=4\nu^2$ must \emph{ always} be present in the spectrum), and therefore the  single-node states are also stable close to the linear limit. For $n=2$ there are two double eigenvalues situated at $\Lambda=4\nu^2$ and $\Lambda=16\nu^2$, and the latter one does split into a complex-conjugate pair, which means that the small-amplitude solutions of this family are unstable. Similar instability also  takes place for families $n=3$ and $n=4$ (we hypothesize that all families with larger $n$ are also unstable near the linear limit). From the perturbation theory it is evident that when the dynamical instability is present, its increment is proportional to $\vep$ and, respectively, proportional to $N^{1/2}$.  This behavior is different from the purely cubic case, where the instability  increment of  small-amplitude modes is proportional to $N$ \cite{AlZez,Alfimov2019}.

We further employ the numerical solution of the eigenvalue problem (\ref{eq:eig})  to address the stability of nonlinear modes of larger amplitude, see Figs.~\ref{fig:families}, Fig.~\ref{fig:energy}(a), and  \ref{fig:global}(a), where solid  and dotted  fragments of plotted curves correspond to stable and unstable nonlinear modes, respectively.   Regarding,  the ground state family  $n=0$ and the single-node family $n=1$, we observe that their solutions remain stable  for modes of  any arbitrary amplitude. This, in particular, means that these families    feature   bistability regions,  where the same   family of nonlinear modes has two stable states with  equal chemical potentials but different numbers of particles. Similar bistability has been earlier encountered for a BEC with spatially inhomogeneous   scattering length  \cite{Zezyulin07} and for  self-sustained \cite{Kaplan} and  guided \cite{Gisin} solitons in the cubic-quintic medium.  We emphasize that in the case at hand the bistability takes place exactly due to the presence of the confining potential, since for zero trapping strength the dependence $\mu(N)$ is monotonous \cite{AstaMalo}, see also the plot $\mu(N)$ for $\nu^2=0$ in Fig.~\ref{fig:families}.

Proceeding to the numerical study of  next families $n=2$ and $n=3$, we confirm that close to the linear limit     nonlinear states are unstable, see the corresponding panels in Fig.~\ref{fig:families}. In the meanwhile, the increase of the number of particles $N$ leads to the stabilization of these families. In terms of the linear stability spectrum, such a stabilization corresponds to a moment when  the complex-conjugate pair of unstable eigenvalues returns to the real axis. The change from instability to stability occurs for the number of particles less than that corresponding to the minimal value of the chemical potential. This means that these families also contain   intervals of bistable chemical potentials, although these intervals are more narrow than those for the  two lowest families with $n=0$ and $n=1$. For larger numbers of particles, families  $n=2$ and $n=3$ have additional finite instability windows which are not shown in Fig.~\ref{fig:families}, but become visible in the more global diagram presented in  Fig.~\ref{fig:global}(a). However, for sufficiently large $N$ these families again become stable, which could be expected from the stability analysis in the TF limit (with purely cubic nonlinearity) performed in \cite{CPK10}. We also notice that in the case at hand   the change of the slope $d\mu / dN$ from negative to positive does not result in the stability change \cite{Yang2012} (as it often happens in other nonlinear wave systems, where the Vakhitov-Kolokolov stability condition \cite{VK} ensures that the solution is unstable when $d\mu/dN>0$).

The fact that the difference $\Delta_n$ between the linear eigenvalue $\tmu_n$ and the minimal chemical potential $\mu_n^*$   weakly depends on the strength of the trapping implies that   the two considered trap strengths correspond to different situations.  Indeed, the chemical potentials of linear eigenstates form an equidistant sequence: $\Delta \mu := \tmu_{n+1} - \tmu_n = 2 \nu$,  and for the strong trap $\nu^2 \gtrsim  1$  the difference between the linear eigenvalues   is much larger than the difference $\Delta_n \approx 0.24$, i.e., $2\nu \gg \Delta_n$.  However, for the weak trap $\nu^2\ll 1$,   the inequality $  2 \nu<\Delta_n$ takes place. In this case   the linear eigenvalue  $\tmu_n$ (which can be considered as a chemical potential  for a gas of noniteracting particles in the harmonic trap)  belongs to the interval of bistable chemical potentials  of    the next  family with $n+1$. This situation is illustrated in Fig.~\ref{fig:energy}(b).

\section{Simulations of dynamics}
\label{sec:dyn}
\begin{figure}
	\begin{center}
		\includegraphics[width=0.999\columnwidth]{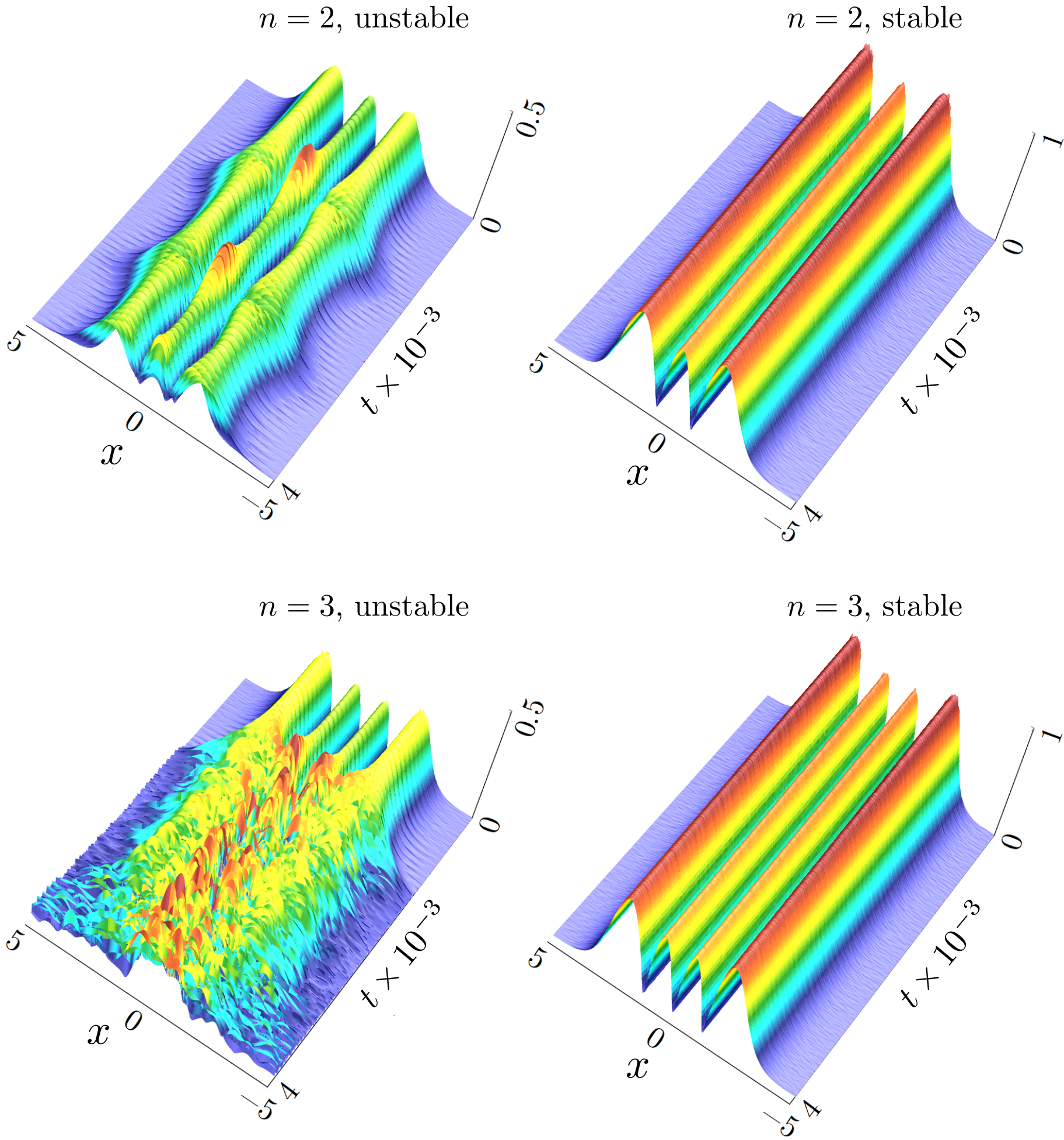}
	\end{center}
	\caption{Plots $|\Psi(x,t)|$  for nonlinear dynamics corresponding to stationary  modes with $n=2$ (upper panels, solutions at $\mu\approx 4.831$) and $n=3$ (lower panels, solutions at $\mu\approx 6.822$). Solutions of smaller amplitudes are unstable, and those with larger amplitudes are stable. \textcolor{black}{Here $\nu^2=1$, and $\sigma_2=\sigma_3=1$.}}
	\label{fig:dyn23}
\end{figure}

Nonlinear dynamics of found stationary states has been simulated by  integrating     the time-dependent GP equation (\ref{GPE}) with a split-step method. To examine the dynamical (bi)stability of stationary states, we solve the initial value problem with the initial condition taken in the form of the stationary wavefunction perturbed by a random noise: $\Psi(t=0,x) = \psi(x)[1 + 0.025(r_1(x) + ir_2(x))]$, where perturbations $r_{1,2}$ are obtained  as  normally distributed pseudorandom numbers. These simulations confirm the existence of bistable states on the lowest ($n=0$) and the first excited ($n=1$) families.   Regarding the next families, $n=2$ and $n=3$, in accordance with the linear stability predictions, we have observed that small-amplitude nonlinear modes are unstable. However, as the amplitude (i.e., number of particles) becomes large enough, the solutions become stable. This difference in stability of excited states of different amplitudes is illustrated in Fig.~\ref{fig:dyn23}, where unstable and stable dynamics are visualized for nonlinear states coexisting at equal values of  the chemical potential.

\begin{figure}
	\begin{center}
		\includegraphics[width=0.999\columnwidth]{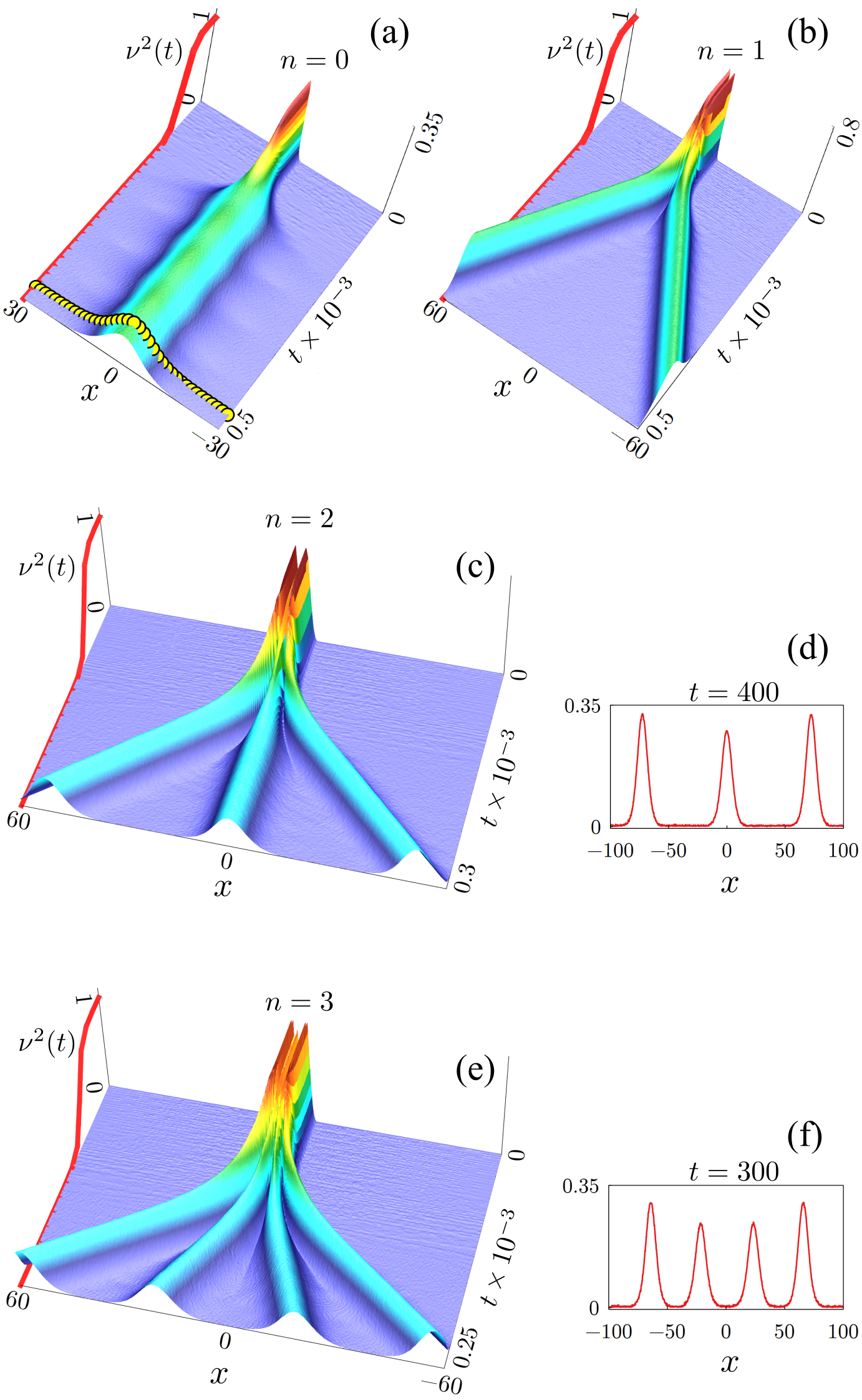}
	\end{center}
	\caption{Plots $|\Psi(x,t)|$  for nonlinear dynamics corresponding to initial condition chosen as a  stationary  mode  with $n=0, 1, \textcolor{black}{ 2, 3}$ 
		as the harmonic trap strength  $\nu^2(t)$ is smoothly decreased from 1 to 0. Dependencies $\nu^2(t)$ are plotted with   bold curves. Dots in (a) show the profile of the solitonlike quantum droplet at $\nu^2=0$ corresponding to the number of particles equal to that of the initial condition;   { (d) and (e) show shapshots of    solutions with $n=2$ and $n=3$ taken at  $t=400$ and $t=300$, respectively.}}
	\label{fig:adia}
\end{figure}

Apart from the direct stability tests, we have addressed several other dynamical scenarios. In particular, we   ran a series of simulations  where the strength of the trapping potential $\nu^2 = \nu^2(t)$  was smoothly decreased down to zero. In this case the nonlinear mode that initially belonged to the ground state family transformed to a solitonlike quantum droplet, while a nonlinear mode from the first family ($n=1$) decoupled in a pair of mutually repulsing droplets of identical form. These results are illustrated in Fig.~\ref{fig:adia}(a) and (b), respectively. \textcolor{black}{Similar behavior can be observed for further families. For instance,  Fig.~\ref{fig:adia}(c) shows how the localized mode from the family $n=2$ breaks into three droplets, one of which is quiescent while two others are moving in opposite directions. As becomes evident from Fig.~\ref{fig:adia}(d), in this case the amplitude of moving droplets is slightly larger than that of the quiescent central droplet.  In a similar way,  the initially trapped localized state from the family $n=3$  fans into four    droplets: two moving to the right with different velocities and two others moving    to the left [see Figs.~\ref{fig:adia}(e-f)].}

\begin{figure}
	\begin{center}
		\includegraphics[width=0.999\columnwidth]{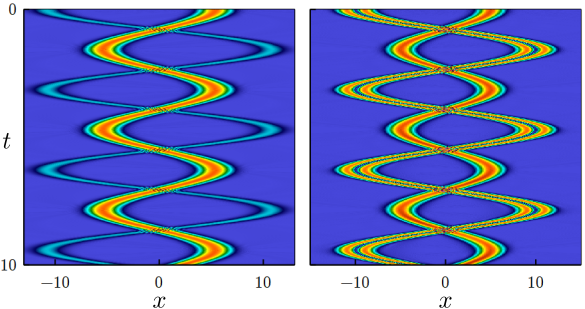}
	\end{center}
	\caption{Pseudocolor plots of composite oscillating solutions composed of two nodeless ground states from the family $n=0$ \textcolor{black}{having equal chemical potential $\mu=0.82$ and different numbers of particles $N\approx 0.15$ and $N\approx 1.32$}  (a), and one ground state with  $n=0$ \textcolor{black}{corresponding to  $\mu=0.82$ and $N\approx 1.32$} and one single-node state with  $n=1$   \textcolor{black}{corresponding to  $\mu=2.8$ and $N\approx 1.57$} (b). \textcolor{black}{In this figure $\nu=1$, $\sigma_2=\sigma_3=1$.}}
	\label{fig:collision}
\end{figure}

It is known that any stationary mode $\psi(x)$ of the GP  equation with the  harmonic potential generates a family of periodically moving solutions given by the explicit formula \cite{exact,Martin} $\Psi(x,t) = \psi(x - X(t)) e^{-i\mu t + i\dot{X}(t) x/2}$, where $X(t)$ is an arbitrary solution of the differential equation $\ddot{X} + 2\nu^2 X=0$ (here dot and double dot  denote first and second derivatives in time $t$). \textcolor{black}{ Clearly, $X(t)$ can be interpreted as a center-of-mass of the oscillating solution.} Exact solutions of this form   enable the systematic investigation of periodically moving quantum droplets. Moreover, assuming that the initial condition is prepared as a superposition of two well-separated quantum droplets: $\Psi(x, t=0) = \psi_1(x - X_1(0))e^{i\dot{X}_1(0)x/2} + \psi_2(x - X_2(0))e^{i\dot{X}_2(0)x/2}$, $|X_1(0) - X_2(0)| \gg 1$, it is possible to simulate  the dynamics corresponding to simultaneous oscillations of two  droplets in the same trap. In Fig.~\ref{fig:collision} we present two examples of   composite oscillating solutions. While preliminary numerical simulations suggest  that the periodic movement robustly persists for indefinite time, an accurate stability study for the oscillating droplets is a  relevant task for future work. \textcolor{black}{For the GP equation with the meanfield cubic nonlinearity such a surprisingly regular   behavior of a pair of colliding harmonically-trapped ground states  has been earlier documented in \cite{Martin} and explained by reducing the GP equation to a dynamical system which treats solitary waves as classical particles and happens to be completely integrable in the   two-particle case.}

\section{Conclusion}
\label{sec:Concl}

 The main goal of our paper has been to develop a systematic analysis of quasi-one-dimensional quantum droplets confined in the parabolic potential. Apart from the fundamental ground states, we have extended the consideration onto the families of nonground excited states which are well-known in the meanfield BEC theory but have  received comparatively little attention in the modified model with the beyond-meanfield corrections. The main results of our study can be summarized as follows.
 
 \begin{itemize}
  \item[(1)] Apart from the family of trapped ground states, there exists a sequence of excited families. Either the ground state family and excited states bifurcate from eigenstates of the underlying quantum harmonic oscillator and feature   nonmonotonous behavior of the chemical potential on the number of particles. Each family has the minimal chemical potential. The difference between the chemical potential in the linear limit and the mimimal chemical potential exhibits a remarkable universality, i.e.,   weakly depends on the family number  and on the strength of the parabolic trap.
  
  \item[(2)] Either the ground state family and excited-state  families  feature a bistability region, where two stable nonlinear modes coexist at the same chemical potential but with different numbers of particles.
  
  \item[(3)] Excited states are unstable close to the linear limit, but become stable as the number of particles increases.  
  
  \item [(4)] In the large-density limit, the trapped states can be described using a modified Thomas-Fermi (TF) distribution which contains larger number of particles  than  the conventional TF cloud in BECs  with purely cubic meanfield repulsion.  
  
  \item[(5)] Smooth  decrease of the harmonic trap strength dynamically transforms the ground state into a solitonlike quantum droplet, while a single-node trapped state transforms into a pair of parting droplets.  More complex trapped states break into multiple droplets.
  
  \item[(6)]  Trapped states   perform stable oscillations around the center of the trap.  A pair of repeatedly colliding trapped states can feature regular dynamics.
  
 \end{itemize}

These results call for a natural generalization onto the effectively multidimensional geometries, where the formation of trapped vortices and vortex rings   and their eventual stabilization using the weak unharmonicity of the trapping potential \cite{unharm,Zezyulin2009} would be a particularly interesting subject. Another   natural extension is to address the role of the confinement  in an essentially two-component asymmetric  mixture, where additional instabilities can emerge \cite{Karta2022}.  \textcolor{black}{Finally, it can be relevant to explore the effect of the external trapping in the modified theory of quantum droplets which accounts for bosonic pairing \cite{pairing1,pairing2} and provides a better agreement between the analytical estimate of the quantum droplet energy and diffusion Monte Carlo simulations \cite{Petrov2016,Parisi1,Parisi2}. The pairing theory is expected to modify  the balance between the competing repulsion and attraction in the corresponding effective one-dimensional Gross-Pitaevskii equation and therefore can lead to  a quantitative change of the results presented herein.}

\appendix

\section{Linear stability analysis}
\label{app:stab}

Using the   standard substitution for the perturbed solution $\Psi = e^{-i\mu t}[\psi(x) + \xi(x,t)]$ and performing the linearization of the GP equation (\ref{GPE}), we find that    the perturbation $\xi(x,t)$ obeys the following equation    (we bear  in mind that wavefunction $\psi$ is real-valued):
\begin{equation}
i\xi_t = - \xi_{xx} - (\mu - \nu^2 x^2)\xi  - \frac{\sigma_2}{2}|\psi|(3\xi + \bar{\xi}) + \sigma_3 \psi^2(2\xi + \bar{ \xi}),
\end{equation}
where $\bar{\xi}$ is the complex-conjugate of $\xi$. 
Separating the perturbation into real and imaginary parts, $\xi = \chi + i\varphi$, we obtain a pair of equations   $\chi_t = - L^-\varphi$, $ \varphi_t = L^+\chi$,
where operators $L^\pm$ are given in Eq.~(\ref{eq:Lm}).
Therefore the stability eigenproblem can be written down as $\Lambda \zeta = L^+L^-\zeta$, where $\Lambda$ is the eigenvalue. The solution is said to be stable if and only if all eigenvalues $\Lambda$ are real and nonnegative. 

Using the pertubation expansions (\ref{eq:expans1})--(\ref{eq:expans2}), for stability of nonlinear modes bifurcating from the $n$th linear eigenstate,  we have $L^+ L^- = \cL_n^2 + \vep M_n + o(\vep)$, where $\cL_n = \partial_x^2 + \tmu_n - \nu^2 x^2$, and 
\begin{equation}
M_n = \cL_n(\mu_n^{(1)} + \sigma_2 |\tpsi_n|) +  (\mu_n^{(1)} + 2\sigma_2 |\tpsi_n|)\cL_n.
\end{equation}
At $\vep=0$ the linear stability operator becomes equal to  $\cL_n^2$.  For each $n=0,1,\ldots$  there are exactly $n$ double (more precisely, semisimple) eigenvalues in  the spectrum: $\Lambda_{n,k} = 4\nu^2(k-n)^2$, where $k=0, 1, \ldots, n-1$ (see \cite{AlZez} for more detailed discussion). As $\vep$ departs   from zero, each double eigenvalue generically splits into a pair of simple eigenvalues. This process can be described using the expansion  $\Lambda_{n,k}^{(1,2)} = 4\nu^2(k-n)^2 + \vep \omega_n^{(1,2)} + o(\vep)$, where coefficients $\omega_n^{(1,2)}$  are   eigenvalues of the $2\times 2$ matrix 
\begin{eqnarray*}
	\tilde{M}_{n,k}=\left(
	\begin{array}{cc}
		\langle M_n\tilde\psi_k,\tilde\psi_k\rangle & \langle
		M_n\tilde\psi_k,\tilde\psi_{2n-k}\rangle
		\\[2mm]
		\langle M_n\tilde\psi_{2n-k},\tilde\psi_k\rangle & \langle
		M_n\tilde\psi_{2n-k},\tilde\psi_{2n-k}\rangle
	\end{array}
	\right).
\end{eqnarray*}
The entries of matrix $	\tilde{M}_{n,k}$ can be represented as
\begin{eqnarray*}
\langle M_n\tilde\psi_k,\tilde\psi_k\rangle = 2\sigma_2(n-k)\int |\tpsi_n|(3\tpsi_k^2 - 2\tpsi_n^2)dx,\\[2mm]
\langle
M_n\tilde\psi_k,\tilde\psi_{2n-k}\rangle = -	\langle M_n\tilde\psi_{2n-k},\tilde\psi_k\rangle  \hspace{2cm} \nonumber\\
=2\sigma_2(n-k)\int |\tpsi_n|\tpsi_k \tpsi_{2n-k}dx,\\
\langle
M_n\tilde\psi_{2n-k},\tilde\psi_{2n-k}\rangle \hspace{4.5cm}  \\ =  -2\sigma_2(n-k)\int |\tpsi_n|(3\tpsi_{2n-k}^2 - 2\tpsi_n^2)dx,
\end{eqnarray*}
where $\int = \int_{-\infty}^\infty$.  Since the eigenfunctions $\tpsi_n(x)$ are available in the explicit form from Eq.~(\ref{eq:twn}), matrices $\tilde{M}_{n,k}$  and their eigenvalues   can be found with a computer algebra software.

\acknowledgments

The work was supported by the Priority 2030 Federal Academic Leadership Program.

\end{document}